\documentclass[preprint2]{aastex62}

\newcommand\aastex{AAS\TeX}

\usepackage{epstopdf}

\received{November 1, 2018}
\revised{November 1, 2018}
\accepted{\today}

\submitjournal{ApJ}

\shorttitle{\aastex\ Blobs in a coronal jet}
\shortauthors{Zhang \& Ni}

\begin{document}

\title{Subarcsecond blobs in flare-related coronal jets}

\correspondingauthor{Q. M. Zhang}
\email{zhangqm@pmo.ac.cn}

\author[0000-0003-4078-2265]{Q. M. Zhang}
\affil{Key Laboratory for Dark Matter and Space Science, Purple Mountain Observatory, CAS, Nanjing 210034, China}

\author{L. Ni}
\affil{Yunnan Observatories, Chinese Academy of Sciences, 396 Yangfangwang, Guandu District, Kunming, 650216, China}
\affil{CAS Key Laboratory of Solar Activity, National Astronomical Observatories, Beijing 100012, China}

\begin{abstract}
In this paper, we report our multiwavelength observations of subarcsecond blobs in coronal jets. 
In AR 12149, a C5.5 circular-ribbon flare occurred at $\sim$04:55 UT on 2014 August 24, which consisted of a discrete circular ribbon and a short inner ribbon inside.
Two jets (jet1 and jet2) were related to the flare. Jet1 appeared first and experienced untwisting motion during its early propagation along a closed coronal loop.
Jet2 appeared 6 minutes later and propagated upward along another closed loop. During its initial phase, a big plasmoid was ejected out of jet2 at a speed of $\sim$150 km s$^{-1}$.
After the flare peak time (05:02 UT), multiple bright and compact blobs appeared in the lower part of jet2, which were observed by the Slit-Jaw Imager (SJI) 
on board the \textit{Interface Region Imaging Spectrograph}. The blobs observed by SJI in 1330 {\AA} have sizes of 0$\farcs$45$-$1$\farcs$35, 
nearly 84\% of which are subarcsecond ($<$1\arcsec). The mean value and standard deviation of the sizes are 0$\farcs$78 and 0$\farcs$19, respectively. 
The velocities of the blobs range from 10 to more than 220 km s$^{-1}$, some of which decelerate and disappear during the upward propagation.
Three of the blobs had their counterparts in extreme-ultraviolet (EUV) wavelengths observed by the Atmospheric Imaging Assembly (AIA) on board the \textit{Solar Dynamics Observatory} spacecraft.
The velocities are almost identical in ultraviolet (UV) and EUV wavelengths. We propose that the blobs observed in 1330 {\AA} are the cool component ($\sim$0.025 MK), 
while the blobs observed in EUV are the hot component of several MK. In jet1, only one blob was present with a size of $\sim$1\arcsec and a velocity of $\sim$40 km s$^{-1}$. 
We conclude that the blobs are created by the tearing-mode instability of the current sheet at the base or inside the coronal jets. 
Our results have important implication for uncovering the fine structures of coronal jets and understanding the relationship between the blobs observed in UV and EUV wavelengths. 
\end{abstract}


\section{Introduction} \label{sec:intro}
Jet-like eruptions are prevalent in the solar atmosphere, such as spicules \citep{dep07}, H$\alpha$ surges \citep{can96}, chromospheric jets \citep{tian14}, and coronal jets \citep{shi92}.
Coronal jets are transient collimated outflows of hot plasma in the solar corona, which are associated with flares or microflares at the base \citep{rao16}.
Coronal jets are observed in the extreme-ultraviolet (EUV) and soft X-ray (SXR) wavelengths \citep{shi96,inn16,zqm17}. According to the different morphology of jets, 
they are previously classified into the anemone jets \citep{yoko96} and two-sided jets \citep{tian17,zheng18}. The temperatures of the jet main body 
are a few MK \citep{shi00}, while the temperatures of the jet base can reach up to tens of MK \citep{bai09}. The electron densities of jets are 0.7$-$4.0$\times$10$^9$ cm$^{-3}$
with a mean value of 1.7$\times$10$^9$ cm$^{-3}$. The apparent widths and heights of jets are 5$-$100 Mm and 10$-$400 Mm, respectively. 
The velocities are 10$-$1000 km s$^{-1}$ with a mean value of 200 km s$^{-1}$ \citep{shi00}. The magnetic field strength of the jets are a few Gauss \citep{puc13}.
The lifetimes of jets range from 10 to 70 minutes, with a median value of 20$-$40 minutes \citep{nist09}. Coronal jets tend to reoccur from the same region, 
which are called homologous jets \citep{mad11,mad12,hua12b,mul17}. Apart from the radial motion, they occasionally experience swirling or untwisting motion \citep{zqm14a}.

It is generally accepted that coronal jets result from impulsive release of magnetic free energy via magnetic reconnection. Numerical simulations and observations 
suggest that several triggering mechanisms of jets are at work, such as magnetic flux emergence \citep{yoko95,more08}, magnetic cancellation \citep{chi08,pan16}, 
mini-filament eruption \citep{ster15,hong16,zhu17,wyp18}, and photospheric rotation \citep{par09}. \citet{moo10} classified the coronal jets into two types, the standard type
and blowout type. The standard jets origin from magnetic reconnection between the emerging magnetic field and the preexisiting coronal field \citep{puc13}, 
while the blowout jets come from the minifilament eruptions \citep{li15,shen17}.

The generation of magnetic islands (or called plasmoids) in a thin current sheet as a result of tearing-mode instability (TMI) (or called plasmoid instability) has been predicted in 
theory \citep{fur63,bis00} and extensively investigated in numerical simulations \citep[e.g.,][]{bha09,shen11,mei12,ni12a}. The creation and cascade of magnetic islands facilitate 
the magnetic reconnection rate and speed up the diffusion of magnetic energy \citep{bha09,hua10,ni15}. Numerical simulations and theoretical analysis have proven that plasmoid 
instability sets in when the Lundquist number ($S$) exceeds the critical value ($S_c=10^{3}$$-$10$^{4}$) \citep{bha09,hua10,hua12a,ni10,ni12b,ni13,com16,hua17}. 
Above the solar photosphere, the Lundquist number is usually 10$^{6}-$10$^{14}$ \citep{pri00}. Therefore, plasmoid instability is expected to occur frequently in the solar eruptions, 
which are related to magnetic reconnection \citep{kar12}. Discrete and intermittent plasmoids are observed in the long current sheet behind coronal 
mass ejections (CMEs) \citep[e.g.,][]{ohy98,ko03,asai04,lin05,tak12,guo13,kum13,cheng18}. In the chromospheric anemone jets, recurring plasmoids with a size of $\sim$0.1 Mm 
are reported by \citet{sin12} and excellently reproduced by the numerical simulations \citep{yang13}. Recently, intermittent plasmoids are detected in ultraviolet (UV) bursts \citep{rv17}. 
The highly dynamic features ($\leq$0$\farcs$2) have line of sight velocities of $\sim$40 km s$^{-1}$ and a lifetime of orders of seconds. The plasmoid cascading theory and numerical 
simulations indicate that the high-order plasmoids with a length scale of ion inertial length of $\sim$1 m can appear in the reconnection process \citep{shi01,ni15,ni18}. 
However, the smaller plasmoids with a length scale of 0$\farcs$1 have not been observed due to the limited spatial resolutions and time cadences of solar telescopes.

For the first time, \citet{zqm14b} reported the discovery of intermittent blobs in recurring coronal jets at the active region (AR) boundary. In a follow-up work, \citet{zqm16a} reported 
the observations of multiple blobs embedded in recurring jets propagating along a large-scale closed loop in the quiet region. The bright and compact features have a size of 
1.5$-$3 Mm, temperature of 1.8$-$3.1 MK, electron density of 1.7$-$2.8$\times$10$^9$ cm$^{-3}$, a lifetime of 24$-$60 s, and a velocity of 120$-$450 km s$^{-1}$.
The authors proposed that the blobs are plasmoids created by TMI of the current sheet and propagate along the jets, which is justified by state-of-the-art 3D numerical 
simulations \citep{wyp16}. \citet{ni17} explored in detail the blob formation and ejection in coronal jets. It is found that, in addition to TMI, the Kelvin-Helmholtz instability (KHI) 
can also create blob-like vortex features in coronal jets when the plasma beta is relatively high \citep{zhao18}.

Using the high-resolution observations obtained from the \textit{Interface Region Imaging Spectrograph} \citep[\textit{IRIS};][]{dep14}, 
\citet{zz17} investigated the bidirectionally moving blobs that originate from the junction between the spire and the arc-base of a jet. 
However, the relationship between the blobs observed in UV and EUV wavelengths is not fully addressed.
In this paper, we report our multiwavelength observations of the subarcsecond blobs in coronal jets associated with a C5.5 circular-ribbon flare (CRF) on 2014 August 24.
As its name implies, the ribbons of CRFs have circular or quasi-circular shapes \citep{zqm15,zqm16b,zqm16c,li18}.
The paper is structured as follows. Data analysis is described in Section~\ref{sec:data}. Results are shown in Section~\ref{sec:result}. 
In Section~\ref{sec:discuss}, we compare our findings with previous works.
In Section~\ref{sec:summary}, we give a brief summary of the results and draw a conclusion.

\section{Instruments and data analysis} \label{sec:data}
The C5.5 flare taking place in NOAA AR 12149 (N10E44) was observed by the Atmospheric Imaging Assembly \citep[AIA;][]{lem12} on board the \textit{Solar Dynamics Observatory} 
\citep[\textit{SDO};][]{pes12} spacecraft. AIA takes full-disk images in UV (1600 and 1700 {\AA}) and EUV (94, 131, 171, 193, 211, 304, and 335 {\AA}) wavelengths. 
The AIA level\_1 data were calibrated using the standard \textit{Solar Software} (\textit{SSW}) programs \textit{aia\_prep.pro}. 
Compared with AIA, the field of view (FOV) of the Slit-Jaw Imager (SJI) on board \textit{IRIS} is much smaller, being 126\arcsec$\times$129\arcsec. 
SJI observed the flare and jets in 1330 {\AA}, which includes two \ion{C}{2} lines at 1335 {\AA} and 1336 {\AA} formed in the upper chromosphere ($T\approx0.025$ MK). 
The \textit{IRIS}/SJI level\_2 data were directly used and coaligned with the AIA images using the cross correlation method.
SXR fluxes of the flare in 0.5$-$4 {\AA} and 1$-$8 {\AA} were recorded by the \textit{GOES} spacecraft.
The observational parameters, including the instrument, wavelength, time cadence, and pixel size are summarized in Table~\ref{tab:para}.

\begin{deluxetable}{ccccc}
\tablecaption{Description of the observational parameters \label{tab:para}}
\tablecolumns{5}
\tablenum{1}
\tablewidth{0pt}
\tablehead{
\colhead{Instru.} &
\colhead{$\lambda$} &
\colhead{Time} & 
\colhead{Caden.} & 
\colhead{Pix. size} \\
\colhead{} & 
\colhead{({\AA})} &
\colhead{(UT)} & 
\colhead{(s)} & 
\colhead{(\arcsec)}
}
\startdata
\textit{IRIS}/SJI & 1330 & 04:30$-$05:22 & $\sim$18.7 & 0.166 \\
\textit{SDO}/AIA & 131$-$335 & 04:30$-$06:00 & 12 & 0.60 \\
\textit{GOES}  & 0.5$-$4.0  & 04:30$-$06:00 & 2.05  & \nodata \\
\textit{GOES}  & 1$-$8  & 04:30$-$06:00 & 2.05  & \nodata \\
\enddata
\end{deluxetable}

\section{Results} \label{sec:result}

\subsection{C5.5 flare and double jets} \label{sec:flare}
\begin{figure}
\plotone{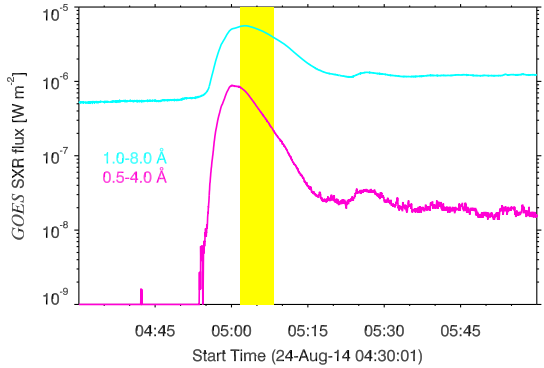}
\caption{SXR light curves of the C5.5 flare on 2014 August 24. 
Cyan and magenta lines represent the light curves in 1$-$8 {\AA} and 0.5$-$4.0 {\AA}, respectively.
The yellow area denote the time of blobs in jet1 and jet2.
\label{fig1}}
\end{figure}

Figure~\ref{fig1} shows the \textit{GOES} SXR light curves of the C5.5 flare. It is clear that the SXR fluxes increase very slowly during 04:30$-$04:55 UT. 
In the impulsive phase, the fluxes rise rapidly to the peak values around 05:02 UT, which is followed by a gradual decay phase until $\sim$05:25 UT. The lifetime of the flare is $\sim$30 minutes.

\begin{figure*}
\plotone{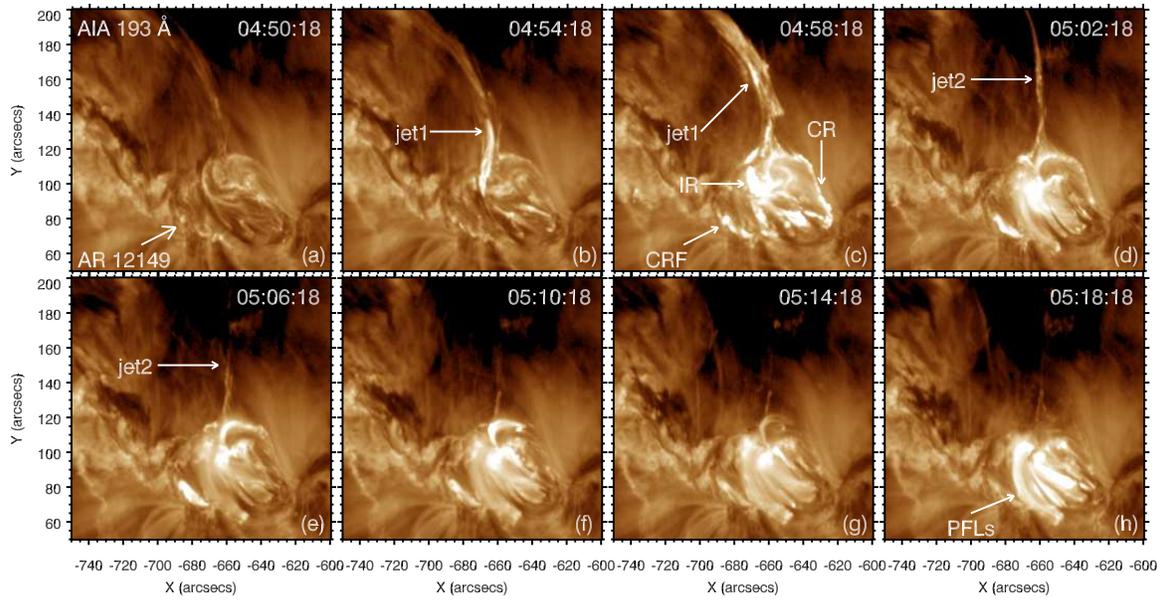}
\caption{Eight snapshots of the AIA images in 193 {\AA} during 04:50$-$05:20 UT. 
The white arrows point to the AR, flare, circular ribbon (CR), inner ribbon (IR), post-flare loops (PFLs), jet1, and jet2.
(An animation of this figure is available.)
\label{fig2}}
\end{figure*}

In Figure~\ref{fig2}, the evolution of the flare is illustrated by eight snapshots of the AIA images in 193 {\AA} (see also the online movie \textit{CRF193.mov}). In the preflare phase before 04:55 UT,
there was no significant brightening in AR 12149 (see panel (a)). At the very beginning of the impulsive phase, a twisted jet (jet1) spurted out of the AR (see panel (b)). About four minutes later,
the jet (jet1) rotated and propagated further and higher along a closed coronal loop in the northeast direction. 
Meanwhile, the EUV intensities of the flare ribbons at the jet base, including the circular ribbon (CR) and inner ribbon (IR), reached their maxima (see panel (c)). The diameter of CR is
$\sim$1$\arcmin$. The short and bright IR is surrounded by the thin and fragmented CR, implying a magnetic null point ($\mathbf{B}=0$) above the CRF and the corresponding fan-spine 
topology \citep[e.g.,][]{zqm12,zqm15}. Interestingly, a second slender jet (jet2) appeared and propagated northwards near the flare peak time (see panel (d)). Like jet1, jet2 propagated 
along a large-scale closed coronal loop (see panel (e)). Afterwards, jet1 and jet2 darkened and disappeared sequentially till 05:15 UT (see panel (g)).
The post-flare loops (PFLs) cooled down and faded out gradually in the decay phase (see panel (h)).

\begin{figure*}
\plotone{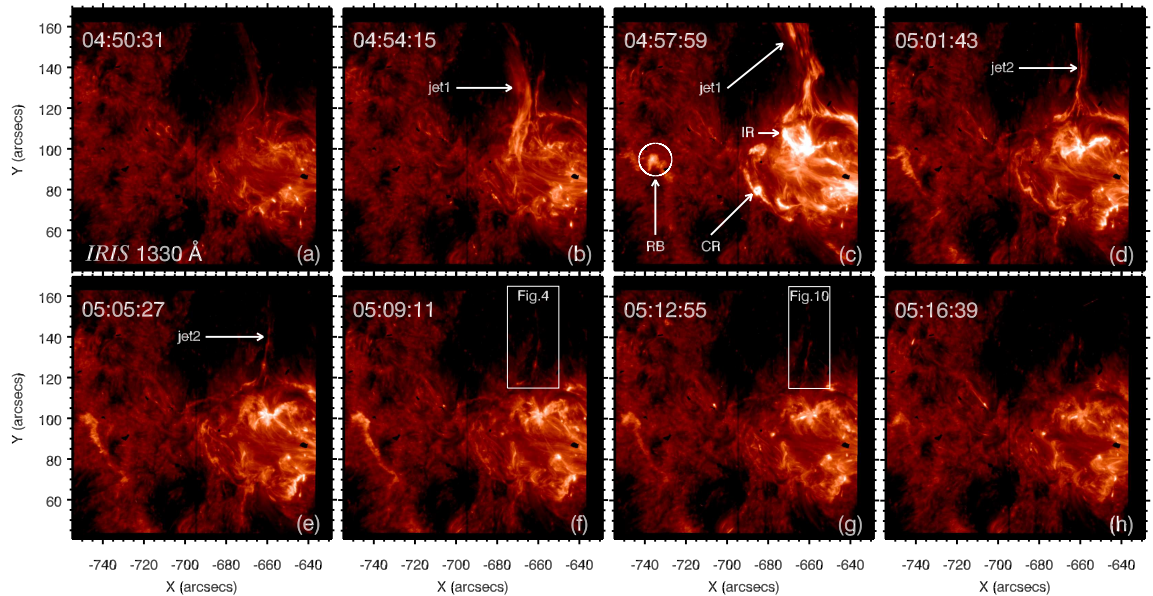}
\caption{Eight snapshots of the \textit{IRIS}/SJI images in 1330 {\AA} during 04:50$-$05:17 UT. 
The white arrows point to the CR, IR, remote brightening (RB), jet1, and jet2. 
The white rectangles in panels (f) and (g) signify the FOVs of Figure~\ref{fig4} and Figure~\ref{fig10}, respectively.
(An animation of this figure is available.)
\label{fig3}}
\end{figure*}

In Figure~\ref{fig3}, the evolution of the flare is illustrated by eight snapshots of the \textit{IRIS}/SJI images in 1330 {\AA} (see also the online movie \textit{CRF1330.mov}).
Due to the limited FOV of SJI, only the eastern part of CR and lower part of the jets are visible (see panel (c)). Like in EUV wavelengths, IR is the brightest feature surrounded 
by the CR. Interestingly, remote brightening (RB) appeared east to the CR, which is enclosed by a small circle (see panel (c)). EUV observations indicate that the RB and CRF is 
connected by a closed coronal loop along which jet1 propagates. Owing to the higher resolution of \textit{IRIS}, the double jets (jet1 and jet2) show more details than those in EUV 
wavelengths.

\subsection{Blobs in jet2} \label{sec:jet2}
\begin{figure*}
\plotone{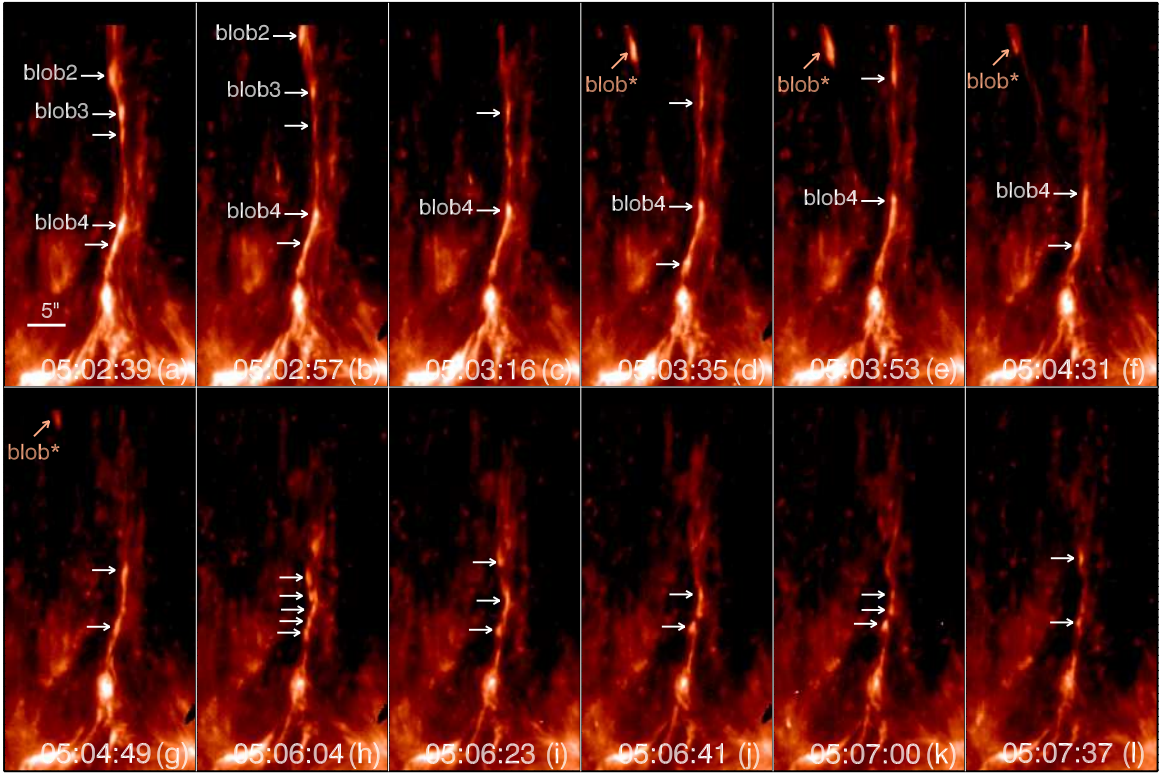}
\caption{Twelve snapshots of the \textit{IRIS}/SJI images in 1330 {\AA} during 05:02$-$05:08 UT. 
The white and orange arrows point to the blobs in jet2 and jet1, respectively.
(An animation of this figure is available.)
\label{fig4}}
\end{figure*}

\begin{figure}
\plotone{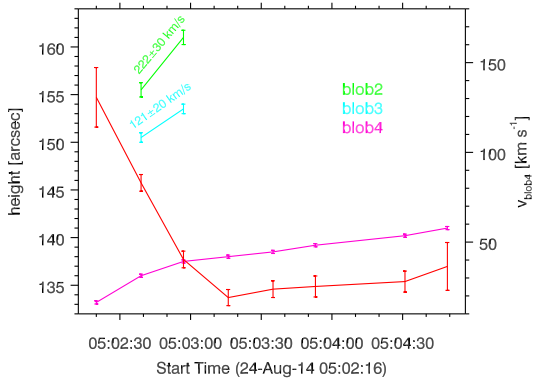}
\caption{Temporal evolutions of the heights of blob2 (green), blob3 (cyan), and blob4 (magenta) observed by \textit{IRIS}.
The temporal evolution of the velocity of blob4 is plotted with a red line.
\label{fig5}}
\end{figure}

Figure~\ref{fig4} demonstrates twelve closeups of jet2 in 1330 {\AA} during 05:02$-$05:08 UT. The FOV (25\arcsec$\times$50\arcsec) is signified by the white rectangle in Figure~\ref{fig3}(f). 
The most prominent features are the tiny, bright, and compact blobs pointed by the white arrows (see also the online movie \textit{blob.mov}). At any time, there are multiple blobs moving upward. 
Some disappeared within a short period of time and new blobs appeared. Three blobs (blob2, blob3, and blob4) with larger sizes are identified and tracked 
by following the gravitational centers of the blobs, which are defined as the intensity-weighted average heights.
The temporal evolutions of the heights are displayed with green, cyan, and magenta lines in Figure~\ref{fig5}. 
Blob2 and blob3 moved so fast that they left the FOV of SJI after 05:03:00 UT. The velocities of blob2 and blob3 are 222$\pm$30 and 121$\pm$20 km s$^{-1}$ in the FOV of SJI.
Blob4 was formed at a relatively lower altitude and moved slower than blob2 and blob3. Therefore, we can track blob4 for a longer time. The temporal evolution of the velocity of blob4 is
plotted with a red line in Figure~\ref{fig5}. It is clear that the velocity decreased quickly from the initial value of $\sim$130 km s$^{-1}$ to 20$-$40 km s$^{-1}$ in the later phase. 

\begin{figure*}
\plotone{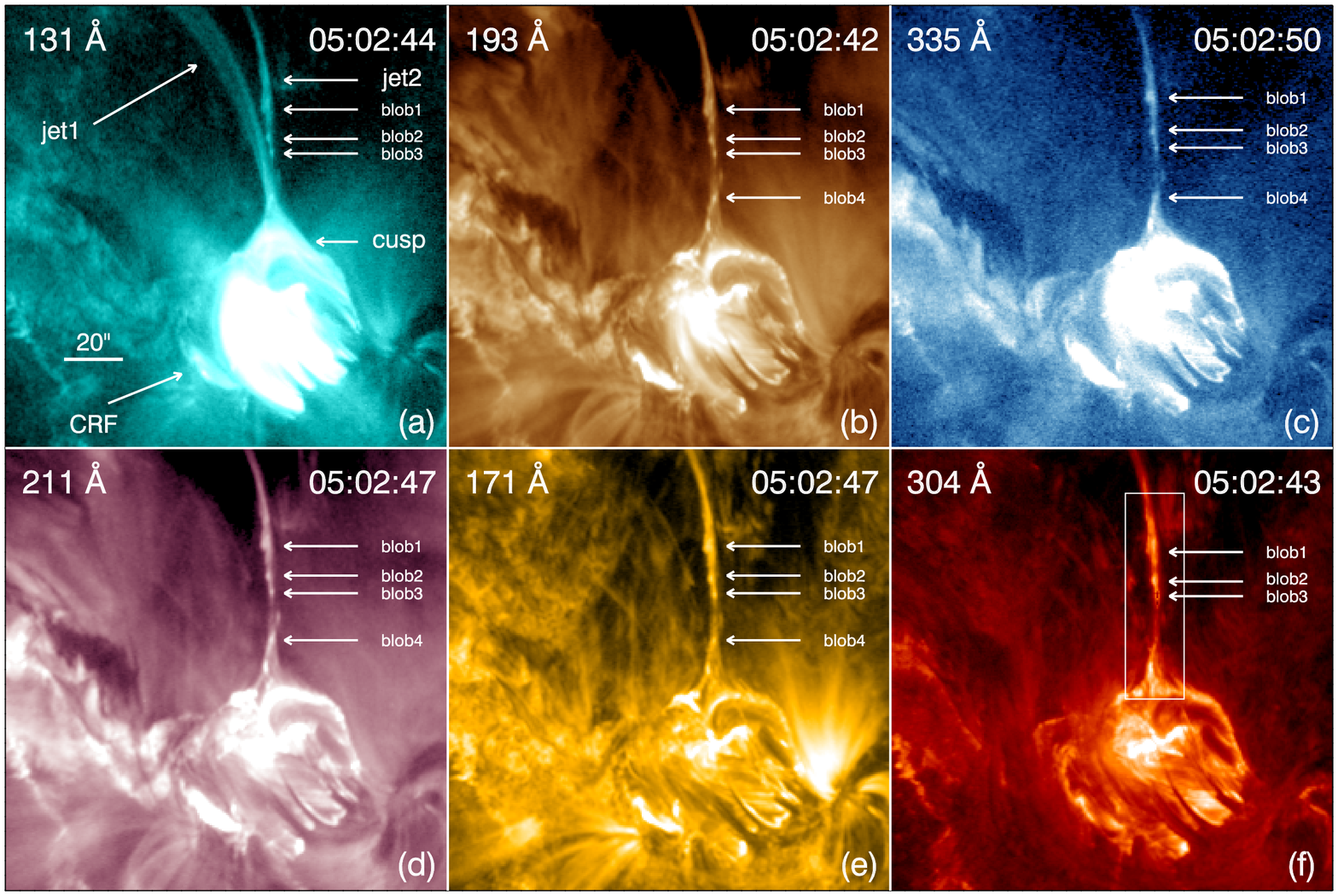}
\caption{AIA EUV images around 05:02:45 UT with the same FOV as Figure~\ref{fig2}. The white arrows point to jet1, jet2, cusp, and multiple blobs in jet2.
In panel (f), the white box signifies the FOV (20\arcsec$\times$70\arcsec) of Figure~\ref{fig7}.
\label{fig6}}
\end{figure*}

Interestingly, we found multiple blobs in jet2 observed by AIA. The EUV images around 05:02:45 UT are displayed in Figure~\ref{fig6}. In each panel, four bright and compact blobs could be
recognized and labeled with blob1, blob2, blob3, and blob4. The diameters of the blobs ($\sim$4$\farcs$14, $\sim$2$\farcs$86, $\sim$2$\farcs$23, and $\sim$2$\farcs$80) are close to the sizes 
reported in previous literatures \citep{zqm14b,zqm16a}. 

\begin{figure*}
\plotone{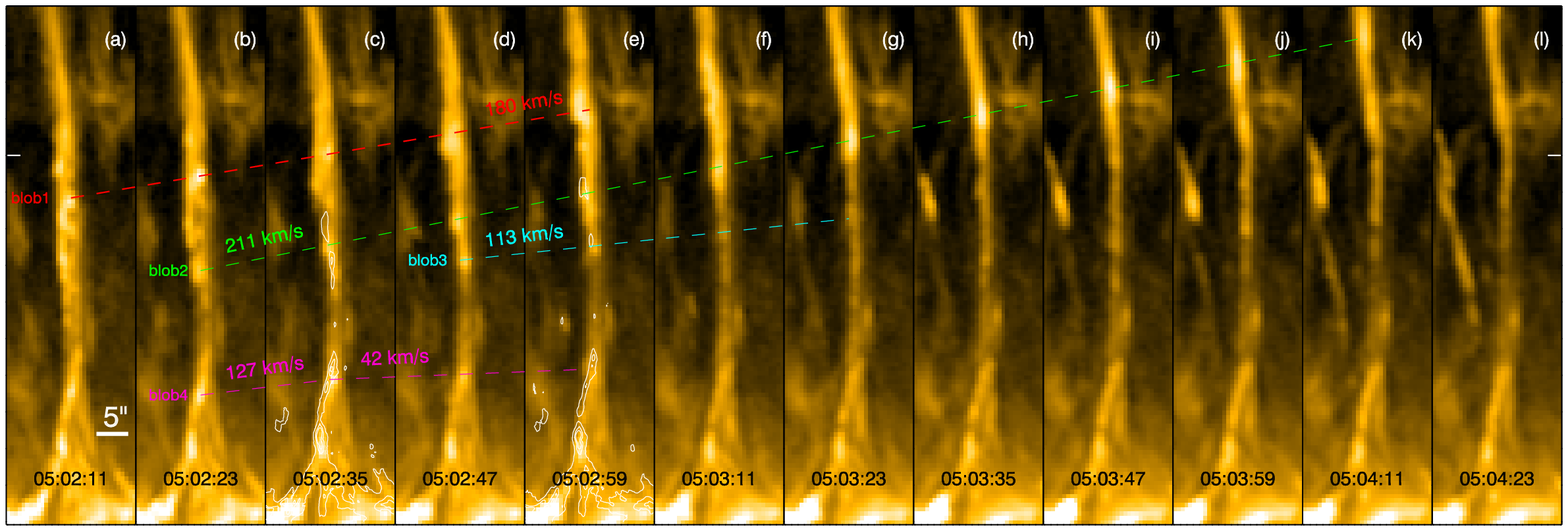}
\caption{Twelve closeups of jet2 observed by AIA in 171 {\AA} during 05:02$-$05:05 UT.
Four blobs (blob1, blob2, blob3, and blob4) are tracked with red, green, cyan, and magenta lines, respectively.
Their linear velocities in the FOV of AIA are labeled. In panels (c) and (e), 
the intensity contours of the 1330 {\AA} images at 05:02:39 UT and 05:02:57 UT are superposed with white lines.
\label{fig7}}
\end{figure*}

In order to track the propagations of the blobs, twelve closeups of jet2 in 171 {\AA} during 05:02$-$05:05 UT are displayed in Figure~\ref{fig7}. It is evident that blob1 was created at the highest 
altitude and moved upward at a speed of 180$\pm$18 km s$^{-1}$. Probably due to the higher temperature and absence of cool plasmas, the formation and propagation of blob1 were not 
captured by SJI. Compared with blob1, blob2 was created at a lower altitude and propagated at a faster speed of 211$\pm$8 km s$^{-1}$. The lifetime of blob2 is close to 2 minutes in EUV 
wavelengths. In panels (c) and (e), we superpose the intensity contours of the 1330 {\AA} images at 05:02:39 UT and 05:02:57 UT with white lines. It is found that blob2 observed in 1330 {\AA} and 
171 {\AA} are cospatial. In other words, blob2 observed in 1330 {\AA} had its counterpart in 171 {\AA}. Blob3 was created around 05:02:47 UT at an altitude slightly higher than blob2 and 
propagated upward at a speed of 113$\pm$12 km s$^{-1}$. The intensity of blob3 decreased too quickly to be recognized after 05:03:23 UT. Blob4 was formed at the lowest altitude. 
The velocity decreased from $\sim$127 km s$^{-1}$ before 05:02:35 UT to $\sim$42 km s$^{-1}$ afterwards. In panel (e), like blob2, blob3 and blob4 observed in 1330 {\AA} 
had their counterparts in 171 {\AA}. Combing with Figure~\ref{fig5}, we can conclude that the velocities of the blobs observed in UV and EUV wavelengths are almost identical.
The coincident presence of blobs in UV and EUV wavelengths indicate that the blobs are multithermal, with their temperatures ranging from 0.025 MK to several MK. 
We have checked the movie of jets in 94 {\AA} and failed to find clear signature of blobs, which means that the temperature of the blobs are lower than the peak temperature 
($\sim$6.3 MK) of 94 {\AA}.

\begin{figure}
\plotone{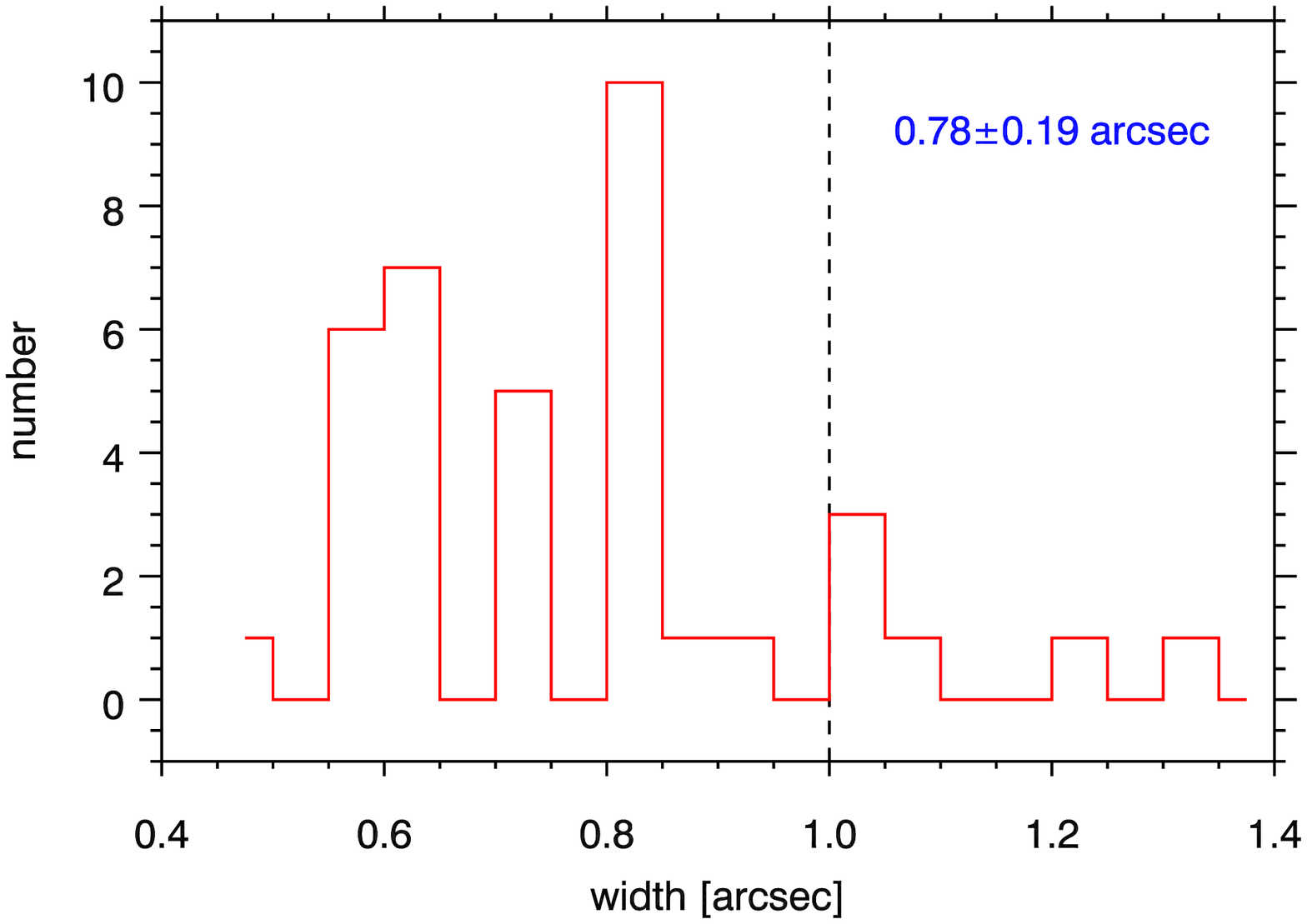}
\caption{Histogram of the widths of the blobs in jet2 observed by \textit{IRIS}/SJI. 
The mean value (0$\farcs$78) and standard deviation (0$\farcs$19) of the widths are labeled. 
\label{fig8}}
\end{figure}

The sizes of the blobs are defined as the full widths at half-maximum (FWHM) of the intensity variation curve along the cuts across the blobs.
We measured the widths of all blobs in jet2 observed by \textit{IRIS} manually and draw a histogram of the widths in Figure~\ref{fig8}. The widths range from 0$\farcs$45 to 1$\farcs$35, 
with a mean value of 0$\farcs$78 and standard deviation of 0$\farcs$19. It is clear that most of the blobs ($\sim$84\%) in jet2 are subarcsecond ($<$1\arcsec) 
and only 16\% have widths greater than 1\arcsec. The sizes of blobs observed by \textit{IRIS} are significantly smaller than those observed by \textit{SDO} \citep{zqm14b,zqm16a,shen17}.

\subsection{Blob in jet1} \label{sec:jet1}
\begin{figure*}
\plotone{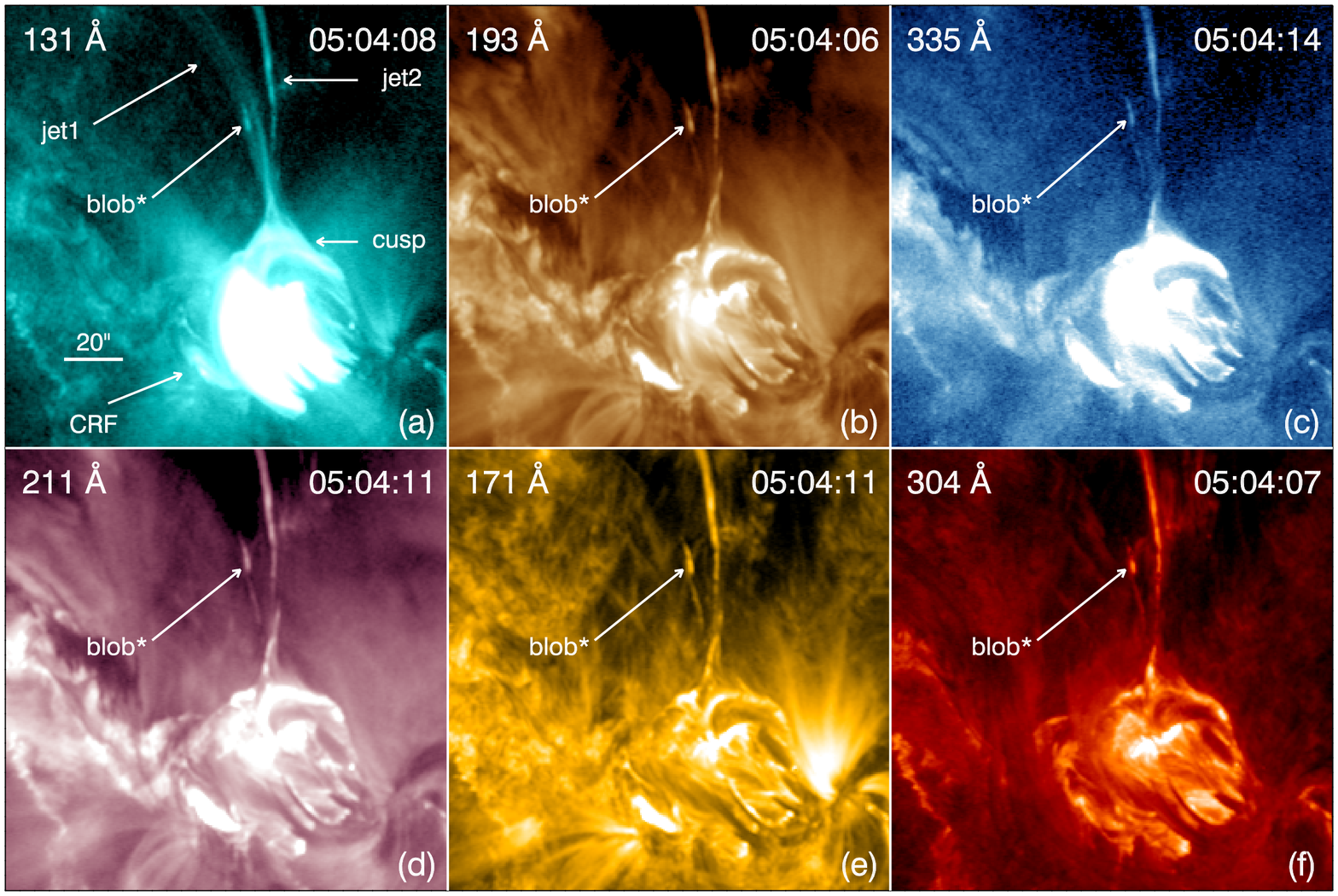}
\caption{AIA EUV images around 05:04:09 UT with the same FOV as Figure~\ref{fig2}. 
The white arrows point to jet1, jet2, cusp, and blob in jet1.
\label{fig9}}
\end{figure*}

Considering the plentiful blobs in jet2, one may also expect a large number of blobs in jet1. We checked carefully the UV and EUV images of the jets, only to find one blob observed by \textit{IRIS} 
and AIA. In Figure~\ref{fig4}(d)-(g), the orange arrows point to the blob in jet1, which is labeled with blob*. The width and velocity of the blob are 1\arcsec$\pm$0$\farcs$2 and 41$\pm$11 km s$^{-1}$. 
In EUV wavelengths, with a size of 2$\farcs$7, blob* appeared at $\sim$05:03:35 UT and propagated upward at a speed of 39$\pm$12 km s$^{-1}$ before disappearing at $\sim$05:04:47 UT.
Figure~\ref{fig9} shows the EUV images observed by AIA around 05:04:09 UT, where blob* in jet1 is pointed by the white arrows. The big difference between the velocities of blob* and the blobs 
in jet2 is probably due to the different environments and velocities of reconnection outflow of the jets. In Table~\ref{tab:blob}, we list the widths and velocities of the blobs observed in 
1330 {\AA} and 171 {\AA}.

\begin{deluxetable}{lcccc}
\tablecaption{Comparison of the widths and velocities of the blobs in 1330 {\AA} and 171 {\AA} \label{tab:blob}}
\tablecolumns{5}
\tablenum{2}
\tablewidth{0pt}
\tablehead{
\colhead{blob} &
\colhead{$w_{1330}$} &
\colhead{$w_{171}$} & 
\colhead{$v_{1330}$} & 
\colhead{$v_{171}$} \\
\colhead{} &
\colhead{(\arcsec)} &
\colhead{(\arcsec)} &
\colhead{(km s$^{-1}$)} &
\colhead{(km s$^{-1}$)}
}
\startdata
blob1 & \nodata & 4.14 & \nodata & 180$\pm$18 \\
blob2 & 1.20 & 2.86 & 222$\pm$30 & 211$\pm$8 \\
blob3 & 0.82  & 2.23 & 121$\pm$20  & 113$\pm$12 \\
blob4 & 0.82  & 2.80 & 20$-$130  & 42$-$127 \\
blob* & 1.0 & 2.70  & 41$\pm$11 & 39$\pm$12 \\
\enddata
\end{deluxetable}

\section{Discussion} \label{sec:discuss}
In the 2D standard flare model, when a filament or flux rope losses equilibrium and erupts into the interplanetary space, a thin current sheet forms between the cusp-shaped PFLs and CME \citep{lin04}. 
The hot and bright current sheet has been frequently 
observed and substantially investigated in EUV and white light wavelengths \citep[e.g.,][]{lin05,sea17}. Observations of multiple blobs in the current sheet have also been reported 
\citep[e.g.,][]{ko03,asai04,tak12,kum13,cheng18}. The sizes and velocities of bidirectional moving blobs are 2\arcsec$-$5\arcsec and hundreds of km s$^{-1}$. The temperatures and densities 
of the blobs are 1.5$-$4 MK and 0.8$-$1.2$\times$10$^{10}$ cm$^{-3}$, respectively. During the magnetic reconnection between a filament and nearby coronal loops, multiple blobs are found 
in the current sheet with temperature of 2.5 MK and electron density of 3.5$-$5$\times$10$^{9}$ cm$^{-3}$ \citep{li16}. In the small-scale chromospheric jets and UV bursts, the blobs are much
smaller \citep{sin12,rv17}.

\begin{figure*}
\plotone{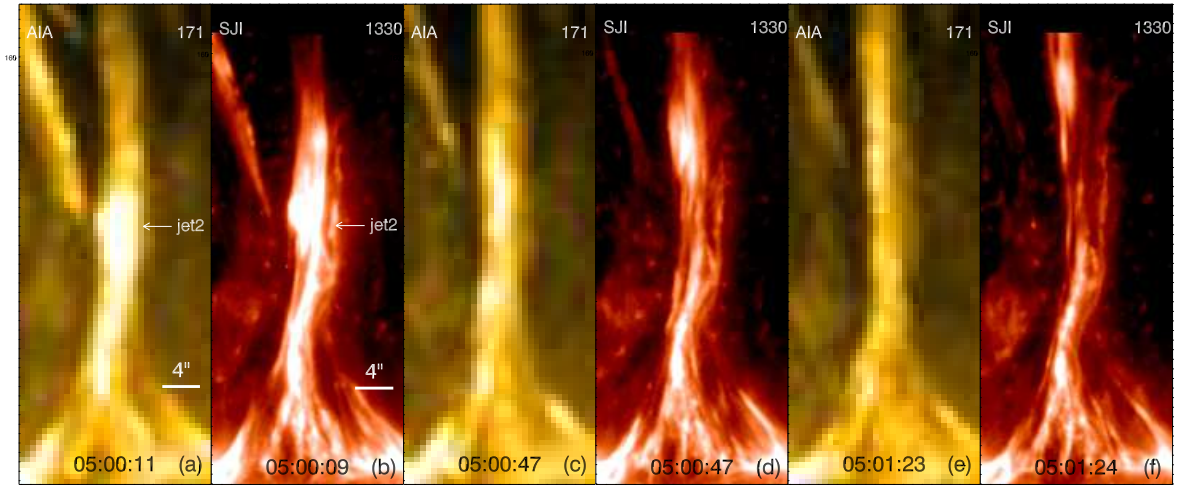}
\caption{Six snapshots of jet2 observed by AIA and SJI during 05:00$-$05:02 UT. 
The FOV (20\arcsec$\times$50\arcsec) is represented by the white rectangle in Figure~\ref{fig3}(g).
\label{fig10}}
\end{figure*}

An open question about the blobs in various scales is that how are they created. It has been predicted that a current sheet may experience TMI when it is thin enough \citep{fur63}. Most of the 
authors attribute the observations of blobs to TMI in a current sheet \citep[e.g.,][]{ohy98,kli00,asai04,kar04,kum13}. The creation, fragmentation, and coalescence of plasmoids in a current sheet 
have been well reproduced in multidimensional numerical simulations \citep{ni12a,ni12b,nis13,yang13,ni15,wyp16}. Based on the numerical simulations of the jet formation due to the magnetic
reconnection between the newly emerging flux and the previous coronal magnetic field, \citet{ni17} proposed that there are two possible mechanisms of the blobs. One is TMI and the other is KHI. 
In this paper, we prefer that the tiny blobs are created by TMI for two reasons. On the one hand, there is no obvious swirling or vortex motion of the blobs as a result of velocity shear in jet2. 
On the other hand, the lower part of jet2 is very close to the CRF (see Figure~\ref{fig3}). A magnetic null point may exist where electric current can accumulate \citep{thu17}. In Figure~\ref{fig10},
six snapshots of jet2 observed by AIA and SJI during 05:00$-$05:02 UT are displayed. At the very beginning of jet2, a ``huge" plasmoid is ejected upward at a speed of $\sim$150 km s$^{-1}$,
which is followed by the formation and ejection of multiple blobs as described above. This is a miniature version of plasmoid eruption and the underlying current sheet in the standard flare 
model \citep[e.g.,][]{shi95,lin04,nis13}.

Some blobs (e.g., blob1) are firstly seen at locations far away from the flaring site or cusp of the jets. These remote blobs could be associated with the shocks as interpreted by \citet{hua18b}.
In the numerical simulations of \citet{ni17}, the bright point pointed by the white arrows are caused by the fast reconnection outflows associated with shocks (see their Fig. 4(a)). 
It is also pointed out that the cusp region at the base of a jet is not the unique region where magnetic reconnections take place. Magnetic reconnection and plasmoid instability also take place 
inside the jet above the cusp region. Hence, the blobs which are firstly seen at locations far away from the flaring site or cusp region are possibly created by the plasmoid instability inside the 
fragmented current sheets along the jet.

In Figure~\ref{fig8}, the sizes of the blobs in jet2 observed by \textit{IRIS}/SJI are 0$\farcs$45$-$1$\farcs$35, with a mean value of 0$\farcs$78 and a standard deviation of 0$\farcs$19. 
For the three blobs (blob2, blob3, and blob4) that are visible in both UV and EUV wavelengths, the sizes in 1330 {\AA} are significantly smaller than those in 171 {\AA} (see Table~\ref{tab:blob}). 
The difference is not caused by the resolution of the instruments since it is beyond the uncertainty brought in by the resolutions.
One possible reason of the difference is that different components of the same blob are separately observed by SJI in 1330 {\AA} and by AIA in 171 {\AA}. Since the formation temperature of 
1330 {\AA} is $\sim$0.025 MK, the blobs observed in 1330 {\AA} represent the cool component of plasmoids. The blobs observed in EUV wavelengths represent the hot component. 
This is in agreement with the previous numerical simulations of TMI in a current sheet \citep{ni16}. The simulations indicate that the plasmas inside the plasmoids can be heated from 
4200 K to $>$10$^5$ K during the magnetic reconnection near the temperature minimum region, and the plasmoids are composed of multithermal plasmas. In our study, we propose 
that the blobs in jet1 and jet2 are plasmoids ejected upward along the jets and some of them are heated to 10$^{5}-$10$^6$ K so that they can be detected in EUV wavelengths. 
\citet{ni17} found that the blobs both inside the current sheet and along the jet are multithermal and have different sizes in different wavelengths. 

Figures~\ref{fig6} and \ref{fig7} reveal that only a few of the blobs created at the base of jet2 can been observed in EUV wavelengths. 
The EUV flux $F_{i}$ at a certain passband (e.g., 171 {\AA}) can be expressed as 
\begin{equation}
F_{i}=\int R_{i}(T)\times \mathrm{DEM}(T)dT\approx R_{i}(T_{max})n_{\mathrm e}^2H,
\end{equation}
where $R_{i}(T)$ is the temperature response function of passband $i$, $\mathrm{DEM}(T)$ stands for the differential emission measure, $T_{max}$ is the electron temperature 
where the response reaches the maximum, $n_{\mathrm e}$ is the electron number density, and $H$ is the integral line-of-sight depth \citep{cheng12}. Therefore, the intensities of blobs depend 
on both density and temperature of plasmas. The EUV intensities of blobs should be high enough to be detected. 
If the blobs have very low density or temperature as a result of insufficient Joule heating, the intensities will be too weak to be detected.

The kinematics of blob4 is interesting. As is shown in Figure~\ref{fig5}, blob4 decelerated from $\sim$130 km s$^{-1}$ at 05:02:20 UT to $\sim$20 km s$^{-1}$ at 05:03:20 UT. 
The deceleration is estimated to be $\sim$1.8 km s$^{-2}$, which is much larger than the gravitational acceleration on the solar surface. There are two possible reasons. The first explanation
is that the flux tubes holding the blobs might be braiding, so that the curvature decelerates the plasma blobs on the plane of sky. 
The original idea of braiding flux tubes where electric current accumulates in the corona was proposed by \citet{par83} and investigated in numerical simulations by \citet{pon17}.
Owing to the extraordinarily high resolutions of spacecrafts, direct observations of braiding fine structures are reported by \citet{cir13} and \citet{hua18a}. 
The second explanation is that the plasmoids in the current sheet might collide and coalesce. As is revealed in the previous numerical simulations \citep[e.g.,][]{ni15,ni16,ni17}, TMI in the
current sheets always result in many reconnection X-points, bidirectional flows, and plasmoids with different sizes and velocities adjacent to the X-points. The collision and coalescence
between these blobs may result in rapid deceleration of the rising blobs.

\section{Summary} \label{sec:summary}
In this paper, we report our multiwavelength observations of subarcsecond blobs on 2014 August 24. The main results are summarized as follows:
\begin{enumerate}
\item{In AR 12149, a C5.5 CRF occurred at $\sim$04:55 UT, which consisted of a discrete circular ribbon with a diameter of $\sim$1$\arcmin$ and a short inner ribbon inside.
Two jets (jet1 and jet2) were related to the flare. Jet1 appeared first and underwent untwisting motion during its early propagation along a large-scale, closed coronal loop. 
Jet2 appeared 6 minutes later and propagated upward along another closed loop. During its initial phase (05:00$-$05:02 UT), 
a big plasmoid was ejected out of jet2 at a speed of $\sim$150 km s$^{-1}$.}
\item{After the flare peak time (05:02 UT), multiple bright and compact blobs appeared in the lower part of jet2, which were observed by \textit{IRIS}/SJI in 1330 {\AA}. 
The sizes of the blobs are 0$\farcs$45$-$1$\farcs$35, with a mean value of 0$\farcs$78 and a standard deviation of 0$\farcs$19. Nearly 84\% of the blobs are subarcsecond, 
and only 16\% are larger than 1\arcsec. The velocities of the blobs range from ten to more than 220 km s$^{-1}$, some of which decelerate and disappear during the upward propagation.
The formation of blobs ended around 05:08 UT.}
\item{Three of the blobs (blob2, blob3, and blob4) in jet2 had their counterparts in EUV wavelengths observed by AIA, and the velocities are almost equal in UV and EUV wavelengths. 
We suppose that the blobs observed in 1330 {\AA} are the cool component ($\sim$0.025 MK),
while the blobs observed in EUV are the hot component of several MK. In jet1, only one blob was clearly observed, with a size of 1\arcsec$\pm$0$\farcs$2 and a velocity of $\sim$40 km s$^{-1}$.}
\item{We conclude that the blobs are created by TMI of the current sheet at the base or inside the coronal jets. Benefiting from the unprecedented spatial resolution of \textit{IRIS}, 
our results are important to uncover the fine structures of coronal jets and address the relationship between the jets observed in UV and EUV wavelengths. 
They will not only impose constraint on the theoretical models of coronal jets, but have implication for magnetic reconnection as well.
Additional case studies and numerical simulations are fairly in demand to figure out the cause and properties of blobs.}
\end{enumerate}

\acknowledgments
We appreciate the referee for valuable and constructive suggestions to improve the quality of this article.
We would also like to thank H. S. Ji, R. Keppens, A. Warmuth, Z. H. Huang, H. Tian, X. L. Yan, Y. D. Shen, and Z. Xu for fruitful discussions. 
\textit{IRIS} is a NASA small explorer mission developed and operated by LMSAL with mission operations executed at NASA Ames Research center 
and major contributions to downlink communications funded by the Norwegian Space Center (NSC, Norway) through an ESA PRODEX contract.
\textit{SDO} is a mission of NASA\rq{}s Living With a Star Program. AIA data are courtesy of the NASA/\textit{SDO} science teams. 
Q.M.Z is supported by the Youth Innovation Promotion Association CAS, NSFC (No. 11333009, 11790302, 11773079, 11729301), the Fund of Jiangsu Province (BK20161618), 
``Strategic Pilot Projects in Space Science'' of CAS (XDA15052200 and XDA15320301), and CAS Key Laboratory of Solar Activity, National Astronomical Observatories (KLSA201716).
L.N. is supported by NSFC (No. 11573064), the Western Light of Chinese Academy of Sciences 2014, the Youth Innovation Promotion Association CAS 2017, 
the Applied Basic Research of Yunnan Province in China (2018FB009), and the key Laboratory of Solar Activity (KLSA201812).

\end{document}